\documentclass[]{llncs}

\usepackage{geometry}
\usepackage[utf8]{inputenc}
\usepackage{caption}
\usepackage{placeins}
\usepackage{amsfonts}
\usepackage{subcaption}
\usepackage{color}
\usepackage{booktabs}
\usepackage[table,xcdraw]{xcolor}
\usepackage{algorithm}
\usepackage{algpseudocode}
\usepackage{graphicx}
\usepackage[htt]{hyphenat}
\usepackage{listings}
\usepackage[normalem]{ulem}
\usepackage{amsmath}
\usepackage{multirow}
\usepackage{multicol}
\usepackage{booktabs}
\definecolor{dkgreen}{rgb}{0,0.6,0}
\definecolor{gray}{rgb}{0.5,0.5,0.5}
\definecolor{mauve}{rgb}{0.58,0,0.82}

\def\bitcoinA{%
  \leavevmode
  \vtop{\offinterlineskip 
    \setbox0=\hbox{B}%
    \setbox2=\hbox to\wd0{\hfil\hskip-.03em
    \vrule height .3ex width .15ex\hskip .08em
    \vrule height .3ex width .15ex\hfil}
    \vbox{\copy2\box0}\box2}}

\title{Union: A Trust-minimized Bridge for Rootstock}

\author{Ramon Amela\inst{1}
\and Shreemoy Mishra\inst{1}
\and Sergio Demian Lerner\inst{1,2}
\and Javier \'Alvarez Cid-Fuentes\inst{1}
}

\institute{RootstockLabs \\
\and Fairgate Labs \\}

\begin{document}
\raggedbottom

\pagestyle{plain}

\maketitle
\begin{abstract}
We present Union, a trust-minimized bridge protocol that enables secure transfer of BTC between Bitcoin and Rootstock.
The growing ecosystem of blockchain systems built around Bitcoin has created a pressing need for secure and efficient bridges to transfer BTC between networks while preserving Bitcoin's security guarantees. 
Union employs a multi-party variant of BitVMX, an \textit{optimistic proving system} on Bitcoin, to create a bridge that operates securely under the assumption that at least one participant remains honest. This $1$-of-$n$ honest approach is strikingly different from the conventional honest-majority assumption adopted by practically all federated systems.
The protocol introduces several innovations: a \textit{packet}-based architecture that allows security bonds to be reused for multiple bridge operations, improving capital efficiency; a system of \textit{enablers} to manage functionaries participation and to enforce penalties; a flexible \textit{light client} framework adaptable to various blockchain architectures; and an efficient \textit{stop watch} mechanism to optimize time-lock management. Union is a practical and scalable solution for Bitcoin interoperability that maintains strong security guarantees and minimizes trust assumptions.
\end{abstract}

\section{Introduction}
\label{sec:intro}

Blockchain interoperability has emerged as a crucial challenge in the cryptocurrency ecosystem, particularly for Bitcoin, the largest and most secure blockchain network. 
While Bitcoin's security and decentralization are unparalleled, its limited programmability has led to the development of numerous secondary blockchain systems that aim to extend its functionality. 
This has created a pressing need for secure and efficient bridges that can transfer Bitcoin's native asset (BTC) between these systems while maintaining Bitcoin's security guarantees.

Traditional bridging solutions often rely on trusted intermediaries or complex multi-signature schemes that introduce significant centralization risks or require substantial collateral. 
These approaches either compromise Bitcoin's trustless nature or prove economically inefficient at scale. 
Furthermore, existing solutions frequently struggle to balance security, capital efficiency, and operational costs.

Current bridges between Bitcoin and a secondary system (typically another blockchain) work by using a \textit{peg-in} transaction to lock some bitcoins into a multisig address controlled by a federation. Once the coins are locked, the secondary system mints an equivalent amount of \textit{synthetic} (or \textit{wrapped}) BTC and transfers it to the user. The challenging part is handling the \textit{peg-out}, where a user wants to convert some synthetic coins back into Bitcoin. This requires the federation to validate the user's request for a peg-out on the secondary system. A majority of the federators must then sign a peg-out transaction on Bitcoin,  transferring BTC to the user. The primary concern with this approach is the need for users to trust that a majority of the federators are honest. Had there been some robust mechanisms to validate the peg-out request transaction on the secondary system directly on Bitcoin - then the trust assumption can be reduced. What is required is a verification system that can validate the outcome of external programs on Bitcoin - a challenging task.

The landscape of Bitcoin verification systems remained relatively stagnant until the release of BitVM~\cite{paper:bitvm}. 
The paper's most significant contribution was not just its technical innovations, but the paradigm shift it inspired. 
Previously, the simplicity of Bitcoin's scripting language led many to believe that all possible advancements had already been made. 
BitVM demonstrated that there was still substantial room for innovation, fundamentally altering the mindset of the entire sector.
The ecosystem quickly adopted these new ideas, with numerous companies and independent developers using these concepts to build general-purpose verifiers on Bitcoin.

Since then, various approaches such as BitVM2~\cite{paper:bitvm2}, BitVMX~\cite{lerner24}, SNARKnado~\cite{paper:snarknado}, and BitSNARK~\cite{paper:bitsnark} have emerged. 
As the universe of Bitcoin verification systems expanded, the next logical phase began.
Building on these verifiers, several new bridging systems have been proposed.
These advancements have led to the development of the BitVM bridge~\cite{paper:bitvm2}, Strata~\cite{paper:strata}, Grail~\cite{paper:bitsnark}, and Clementine~\cite{repo:clementine}.
This paper contributes to this new phase of bridging systems with Union, a trust-minimized bridge protocol that enables the secure transfer of BTC between Bitcoin and Rootstock~\cite{lerner22}, the longest running and leading Bitcoin sidechain.

The key contributions of this paper are: (1) a novel bridge architecture that requires only one honest functionary rather than an honest majority, significantly reducing trust assumptions compared to traditional approaches; (2) an innovative packet-based system that allows multiple bridge operations to share security deposits, improving capital efficiency; (3) a flexible enabler mechanism that effectively manages functionary participation and enforces penalties for misbehavior; (4) an efficient stop watch mechanism that optimizes time-lock management in multi-step verification processes; and (5) a generalized framework for implementing trust-minimized bridges that can adapt to various blockchain architectures while preserving Bitcoin's security properties.

This paper is organized as follows. 
Section~\ref{sec:model} presents the model overview, including the protocol assumptions and goals. 
Section~\ref{sec:union} introduces the Union protocol, detailing the multi-party BitVMX proving system, the peg-in and peg-out processes, security deposits, and the innovative packet and enabler systems. 
Section~\ref{sec:analysis} provides a comprehensive analysis of key design decisions and trade-offs, examining the size of the functionaries set, security deposit calculations, capital efficiency, and parallelism considerations. 
Finally, Section~\ref{sec:conclusion} concludes the paper by summarizing our contributions and findings.

\section{Model Overview}
\label{sec:model}

Blockchain interoperability, including asset bridging, has received considerable attention in the scientific literature~\cite{belchior21,zamyatin21}. The general model for asset bridging can be summarized as follows~\cite{buterin16}.

Let $B_s$ and $B_t$ be two blockchain systems (source and target), and let $C_s$ be a native asset in $B_s$. Bridging $C_s$ from $B_s$ to $B_t$ consists of the following steps: (i) locking (or making unavailable) on $B_s$ an amount $A$ of $C_s$, (ii) communicating and validating on $B_t$ the locking operation on $B_s$, and (iii) unlocking (or making available) an amount $A$ of an asset $C_t$ in $B_t$. This process is usually known as \emph{peg-in}, and the asset $C_t$ is sometimes referred to as a ``wrapped asset''. The process of returning the asset $C_t$ from $B_t$ to $B_s$ is known as \emph{peg-out} and consists of the same steps but reversed.

The security of the bridging process lies in guaranteeing that $C_t$ cannot be unlocked unless the equivalent amount of $C_s$ has been locked previously and vice versa~\cite{paper:sok}. This requires a robust locking mechanism on both blockchain systems, as well as a reliable mechanism to validate information on each of the blockchains that comes from the other system.

\subsection{Protocol assumptions}

The Union protocol is an instantiation of the aforementioned general bridging model with the following components:

\begin{itemize}
    \item \emph{Bitcoin:} Bitcoin is the source blockchain ($B_s$) of the protocol and BTC is the transferred asset ($C_s$).
    \item \emph{Rootstock:} the protocol connects Bitcoin to Rootstock. 
    \item \emph{Multi-party proving system:} the protocol uses BitVMX~\cite{lerner24} in a multi-party setting as a mechanism to optimistically verify computations on Bitcoin. Such verification can be used to restrict the spending of a UTXO based on events external to Bitcoin and by emulating covenant-like behavior. For ease of exposition, we refer to a UTXO secured by a BitVMX setup as a VMXO. The following section provides an overview of BitVMX.
    \item \emph{Functionaries:} the protocol defines a set of entities called functionaries that operate the bridge and participate in the multi-party proving system. Each functionary is identified with their Bitcoin public key as well as their public key (or address) on $S$.
\end{itemize}

The Union protocol is a bridge between Bitcoin and Rootstock. Nevertheless, the Union protocol can be generalized to any secondary blockchain system $S$ under the following assumptions:

\begin{itemize}
    \item \emph{Blockchain security:} both Bitcoin and the secondary system $S$ are secure.
    \item \emph{Blockchain neutrality:} block producers in both Bitcoin and $S$ do not collude with each other or with the functionaries.
    \item \emph{Eventual delivery:} there is a time bound in message processing and blockchain inclusion in both Bitcoin and $S$. This assumption is useful when thinking about the protocol's security from transaction censoring attacks.
    \item \emph{Blockchain relays:} the blockchain system $S$ implements a Bitcoin \emph{relay}~\cite{buterin16}, and thus is able to validate transaction inclusion on Bitcoin. In addition, a light client for $S$ can be implemented such that it can decide on a canonical chain given two competing forks.
    \item \emph{Proving system security:} the BitVMX proving system provides the means to create a Bitcoin UTXO that can be spent \emph{if and only if} some computation is optimistically validated on-chain.
    \item \emph{Functionary set management:} System $S$ has a (ideally consensus-dependent) \textit{registration} mechanism to modify the membership of the functionary set.
    \item \emph{Functionary honesty:} at least one of the functionaries is honest.
\end{itemize}

\subsection{Protocol goals}

We define the following goals when designing the Union protocol:

\begin{itemize}
  \item \emph{Adequate incentives:} functionaries are economically incentivized to operate the protocol, and are economically disincentivized to misbehave.
  \item \emph{Functionary penalization:} there is a mechanism to prevent misbehaving functionaries from participating in the protocol.
  \item \emph{Capital efficiency:} the capital required to disincentivize functionary misbehavior grows sub-linearly with respect to the amount of value locked in the protocol.         
  \item \emph{Liveness:} as long as users lock an equivalent amount of assets on one blockchain (Bitcoin or $S$), they are guaranteed to receive the corresponding assets on the other blockchain.
  \item \emph{Generality:} the protocol is designed to be flexible and can be adapted to various blockchain systems and consensus mechanisms.
\end{itemize}

\section{The Union Protocol}
\label{sec:union}

Since Union relies on BitVMX, we present a brief overview of how BitVMX works in a multi-party environment.

\subsection{Multi-party BitVMX proving system}

BitVMX~\cite{lerner24} is an optimistic proving system that allows validating arbitrary computations on Bitcoin. Using BitVMX, two parties can create a dispute resolution game on-chain, where they take the respective roles of prover and verifier. The prover can then submit proof for some arbitrary computation, while the verifier can challenge the proof in case of disagreement. This proof consists of the input and final state of a virtual CPU after executing a pre-agreed program. To ensure that both parties commit to using the BitVMX mechanism, they must both lock some bitcoin as part of the setup process, creating a UTXO whose spending is restricted by the outcome of the protocol. We refer to such UTXOs as VMXOs.

The dispute resolution game consists of multiple on-chain rounds where the verifier challenges specific parts of the data that the prover has provided, and the prover must respond with the requested information. The game is designed such that if the proof is invalid, the prover will be forced to commit to conflicting data across different rounds. This results in the prover being unable to respond to at least one challenge, as no valid transaction could be created satisfying the constraints established by the previous transaction scripts.  Each round has an associated time lock, and if either party fails to respond within this timeframe, they lose the game and the other party can claim the \textit{reward} (bitcoin locked in the VMXO). The system is optimistic because the prover wins after a predefined time period if the verifier does not challenge the initial proof. The dispute resolution game is implemented as a directed acyclic graph (DAG) of interconnected presigned transactions where BitVMX emulates covenants, using a message-linking scheme to ensure that both parties are obligated to engage in the game once the initial setup transaction is confirmed. 

\begin{figure}[b]
  \centering
  \includegraphics[width=\textwidth]{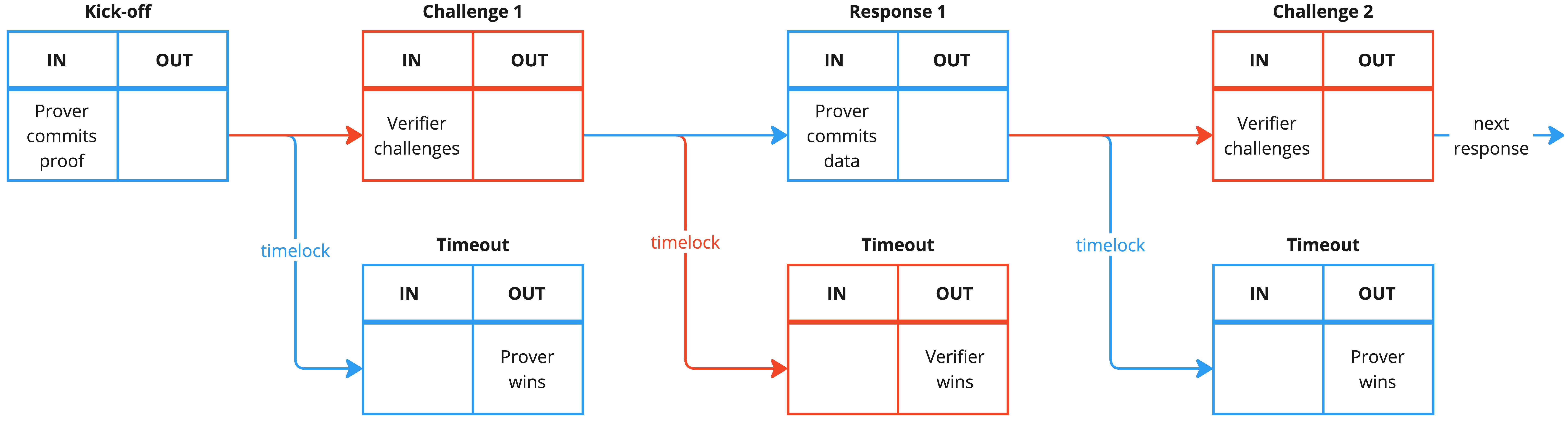}
  \caption{BitVMX Dispute Resolution Channel. The prover submits a proof of a computation that a verifier can challenge through a dispute resolution game. Colors blue and red in the diagram represent transactions from the prover and the verifier respectively.}
  \label{fig:dispute}
\end{figure}

Figure~\ref{fig:dispute} shows a partial illustration of the BitVMX dispute resolution transaction graph, which is a core component of the proving system. We use the abstraction of a communication channel and refer to this DAG component as a \textit{Dispute Resolution Channel}. The figure illustrates a repeated, continuing game between the prover and the verifier which can unravel over several rounds. The bound on the number of rounds will depend on the nature of the computation to be verified and how the protocol is implemented. Each of the leaf nodes in this DAG represents a situation where one party wins and can spend funds locked in the VMXO. To keep the illustration simple, a setup transaction that creates the VMXO from initial deposits, and the transaction where it is spent by the ultimate `winner' are not shown in Figure~\ref{fig:dispute}.

While BitVMX was originally designed for two parties, it can be extended to a multi-party setting where $N$ parties participate in the dispute resolution game. Instead of having a single prover-verifier pair pre-signing one set of transactions, each party creates transaction sets with every other party, allowing them to act as either prover or verifier. This creates a set of DAGs where every participant maintains bilateral dispute resolution channels with all other participants. This arrangement is achieved by having all parties collectively cosign the necessary transaction sets. This establishes a web of interconnected dispute resolution games that maintains the security properties of the original two-party system while distributing the verification responsibilities across multiple participants. After signing all the transactions, all honest parties are expected to delete their private keys in order to guarantee that the only possible spending paths are those that go through the transactions created in the BitVMX setup. However, the system is secure as long as at least one party does not share their private key. The most secure way to ensure it is to actually delete it.

\begin{figure}[b!]
  \centering
  \includegraphics[width=0.7\textwidth]{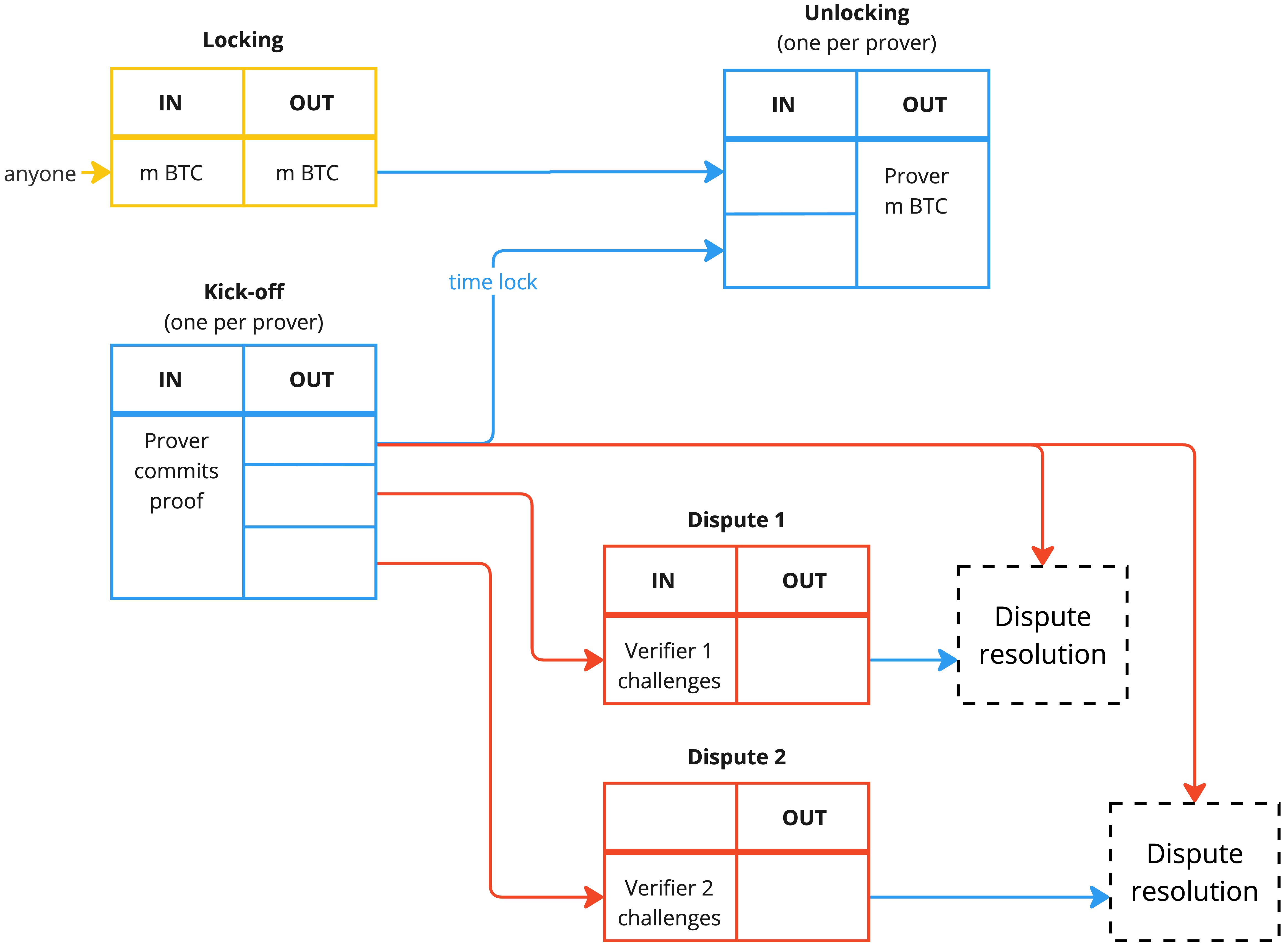}
  \caption{Transaction graph structure for multi-party BitVMX showing the \textit{Locking}, \textit{Kick-off}, dispute, and \textit{Unlocking} transactions. The dispute resolution component corresponds to the channel scheme depicted in Figure~\ref{fig:dispute}.}
  \label{fig:multiparty}
\end{figure}

We can use this multi-party BitVMX scheme to lock BTC in a way that can only be unlocked by providing a valid proof of computation. 
Figure~\ref{fig:multiparty} shows a diagram of this scheme for a use case where a party is willing to pay a set of competing service providers to execute a program and have the output verified optimistically on Bitcoin. 
A single \textit{Locking} transaction pays $m$ BTC that any prover can retrieve by publishing a valid proof of computation in a \textit{Kick-off} transaction. 

Each prover's \textit{Kick-off} transaction contains one output per potential verifier, which allows any party to challenge the proof. 
If the prover loses any dispute resolution process, one of the verifiers spends the \textit{Kick-off} transaction output required to mine the \textit{Unlocking} transaction, thus preventing the unlocking of the BTC. 
For a single \textit{Locking} transaction, the $N$ parties collectively sign $N$ sets of the scheme. 
In this way, any of the parties can act as prover to try and collect the locked BTC, while any party can serve as a verifier and challenge fraudulent attempts. 
For each \textit{Locking} transaction, there are $N$ \textit{Kick-off} and \textit{Unlocking} transactions, and each \textit{Kick-off} transaction connects to $(N-1)$ dispute resolution channel, one for each potential verifier. 
Ultimately, only one of the provers is able to collect the funds by spending the \textit{Locking} transaction output.

This same abstraction can be used for use cases of specific interest like the optimistic verification of zero-knowledge proofs on Bitcoin and to create bridge protocols such as Union.

\subsection{Union bridge protocol overview}
The Union protocol enables trust-minimized transfer of BTC between Bitcoin and Rootstock. The protocol is operated by a set of functionaries who collectively manage a multi-party BitVMX setup on Bitcoin to secure the locked funds. Every peg-in locks some BTC inside a UTXO, each of which is secured by its own BitVMX setup. As we will see, separate BitVMX setups can share and reuse some components of the transaction graph. For now, we note that each peg-in creates a VMXO (a BitVMX secured UTXO). And ultimately, each peg-out will be sourced from a pre-existing VMXO.

The protocol operates under the assumption that at least one of $N$ functionaries is honest. When users want to transfer BTC to Rootstock (peg-in), they lock their funds in the BitVMX setup and receive a wrapped version of BTC on Rootstock. For transfers back to Bitcoin (peg-out), any functionary can act as a prover by advancing the BTC to the user and then recovering these funds from the BitVMX setup by proving the validity of the peg-out operation on Rootstock and the transfer of the correct amount to the user on Bitcoin. The remaining functionaries act as verifiers in this process, ensuring the correctness of the proof. This arrangement allows any functionary to facilitate transfers while being verified by their peers, creating a flexible and secure bridge between the two blockchain systems.

Since the trust assumption of the protocol is that at least one functionary is honest, the size of the functionary set does not negatively impact the security of the protocol. This means that the functionary set can be extended to a large number with minimal requirements for joining (e.g., locking some security deposit). From the point of view of an entity that participates in the functionary set, the protocol is completely trustless.

\subsection{Cross-chain validation}
\label{sec:validation}

For the Union protocol to transfer BTC between Bitcoin and a generic secondary blockchain system $S$ (e.g., Rootstock), it is crucial to prove in one blockchain that specific events have occurred on the other blockchain. This cross-chain validation is achieved through light-client implementations. In the case of peg-ins, we need to prove that a valid deposit transaction exists on Bitcoin. This is possible through $S$'s consensus because, as assumed in Section~\ref{sec:model}, $S$ implements a Bitcoin relay, and thus has the ability to validate a Bitcoin transaction inclusion. In the case of peg-outs, we need to prove on Bitcoin that a valid matching peg-out transaction exists on $S$. The Union protocol assumes that it is possible to implement a light client for $S$, which can then be validated using BitVMX. Both of these assumptions are valid for the specific case of the Rootstock sidechain.

The light client for system $S$ takes a sequence of blocks and verifies that is the canonical chain (as defined by system $S$'s consensus rules). It then verifies whether a valid unlock transaction exists within this sequence. More precisely, the light client for $S$ is a function:
\begin{equation*}
  \texttt{checkChain}(\mathcal{H}_1,P_i,P_o,D_1) ,
\end{equation*}
where
\begin{itemize}
  \item $\mathcal{H}_1=\{H_i,H_{i+1}...,H_{i+k}\}$ is a sequence of block headers in the blockchain $S$
  \item $P_i=(T_i,H_i)$ is an inclusion proof of a peg-in transaction $T_i$ in block $H_i$ in the Bitcoin blockchain
  \item $P_o=(T_o,H_o)$ is an inclusion proof of a peg-out transaction $T_o$ in a block of the blockchain $S$ $H_o$ part of $\mathcal{H}_1$
  \item $D_1$ the total accumulated difficulty of $\mathcal{H}_1$
\end{itemize}

The \texttt{checkChain} function returns true if $\mathcal{H}_1$ is a valid chain of blocks in $S$ that connects a valid peg-in transaction with the corresponding peg-out transaction. In the \textit{Kick-off} transaction, the prover commits to the final state and inputs of \texttt{checkChain}, which serves as proof of computation in order to unlock the BTC unless one of the verifiers triggers BitVMX's dispute resolution. 

However, merely proving the existence of a block sequence and peg-out transaction is insufficient, as a malicious prover could mine a separate sequence of blocks independently from $S$'s canonical chain and present this counterfeit sequence. To prevent this attack, the BitVMX dispute resolution game is modified to allow verifiers to submit a counter-proof consisting of an alternative sequence of blocks. This modification adds a new challenge type to the dispute resolution game shown in Figure~\ref{fig:dispute}, where the verifier presents proof of an alternative block sequence with higher accumulated difficulty than the prover's initial sequence. More precisely, the verifier submits proof of the function
\begin{equation*}
  \texttt{checkAltChain}(\mathcal{H}_2,P_i,P_o,D_2) ,
\end{equation*}
where $\mathcal{H}_2=\{H_i,...,H_{i+m}\}$ is an alternative sequence of block headers with total accumulated difficulty $D_2$. This function returns true if $\mathcal{H}_2$ is a valid chain of blocks that contains $T_i$, but \textbf{does not} contain the block $H_o$. The challenge transaction checks that $D_2>D_1$, and thus the validity of \texttt{checkAltChain} is sufficient to prove that the original sequence of blocks is not part of $S$'s canonical chain.

\begin{figure}[h]
    \centering
    \includegraphics[width=0.75\textwidth]{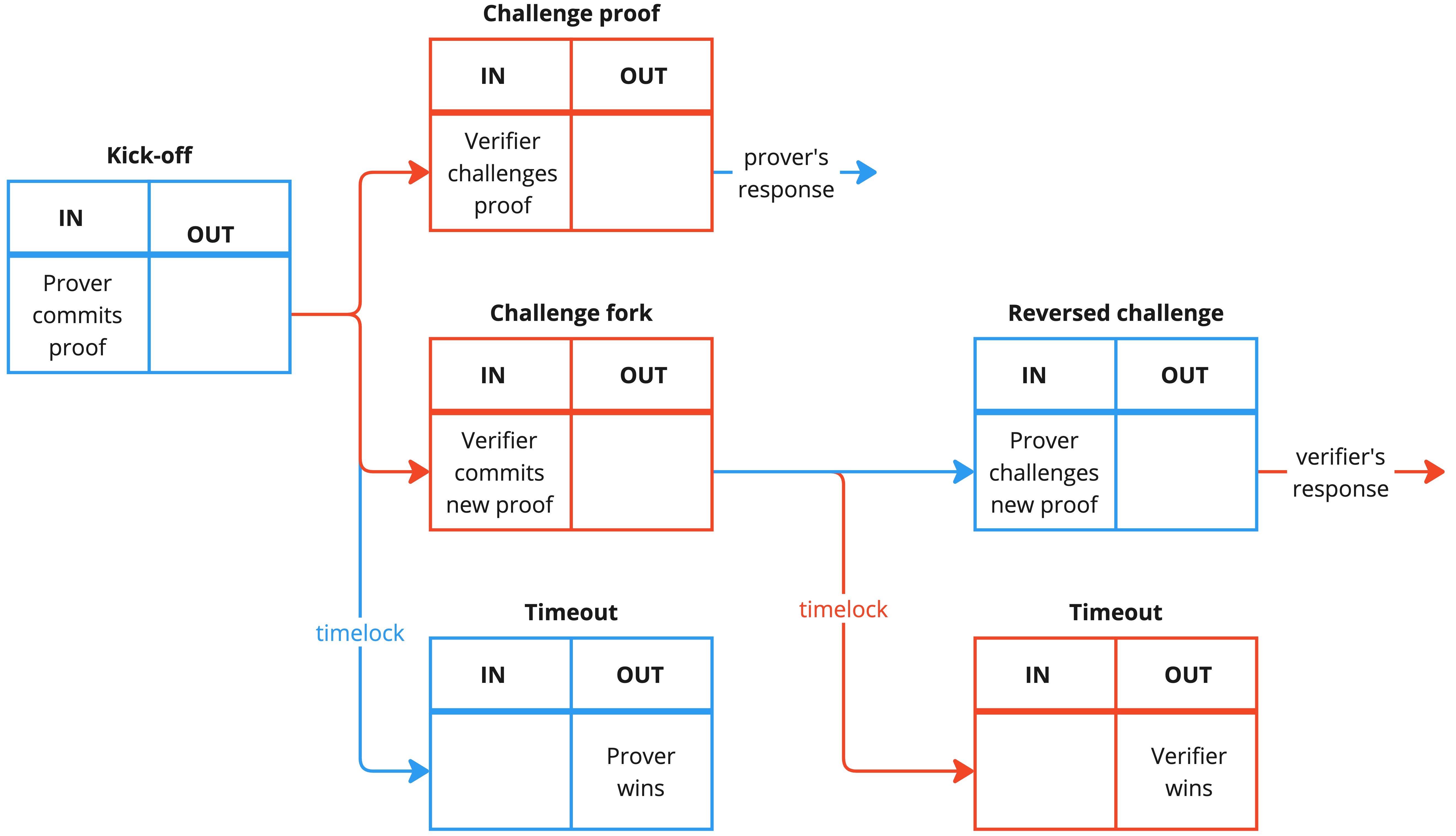}
    \caption{Transaction graph structure of the dispute resolution game with verifier counter-proof using nested BitVMX.}
    \label{fig:counter}
  \end{figure}

The original prover can then challenge this new proof of computation following the same BitVMX dispute resolution mechanism as before, essentially creating a nested game with reversed roles. This adds only one extra step to the original BitVMX protocol as just one of the two proofs of computation needs to be disproved. Figure~\ref{fig:counter} shows a diagram of the modified BitVMX dispute resolution process. The verifier can choose between challenging the prover's computation or the sequence of blocks submitted by the prover.

This counter-proof mechanism ensures that provers can only successfully claim funds by providing proof of transactions that exist in $S$'s canonical chain, as any attempt to use a forked or counterfeit chain can be challenged by presenting the true canonical chain with higher accumulated difficulty.

Given that the maximum size of a Bitcoin block is 4MB, it is impractical to commit the entire sequence of block headers on-chain due to space constraints. 
Even if it were possible to fit the full header sequence within a block, the cost of committing such a large amount of data would be prohibitively high. 
Additionally, we must also commit data, to prove transaction inclusion, further increasing the data requirements.
SNARGs~\cite{gentry2011separating} allow us to present a proof of the correct result of the \texttt{checkChain} function computation without performing the computation itself.
This is particularly advantageous as it eliminates the need to commit to the entire sequence of blocks.
As a result, the input size and the amount of information committed on-chain can be significantly reduced.

\subsection{Peg-in process}

The peg-in process, which enables users to transfer BTC from Bitcoin to Rootstock, uses the same multi-party BitVMX scheme depicted in Figure~\ref{fig:multiparty}. 
The user deposits BTC using the \textit{Locking} transaction shown in the diagram, and this triggers the unlocking of the wrapped BTC on Rootstock. 
Peg-ins are a permissioned process as the multi-party BitVMX setup requires all the functionaries to sign the transactions and delete their private keys. 
Rootstock accepts a deposit transaction as a valid peg-in if it is signed by all the functioanries and confirmed on Bitcoin. 
This is in line with the 1-of-$n$ honest assumption of the Union protocol. 
The peg-in process consists of the following steps:

\begin{enumerate}
    \item The user requests a deposit address from the functionaries for a specific amount of BTC. As will become clear later, the functionary set may restrict peg-in deposits to a predetermined set of values.
    
    \item The functionary set provides the user with a deposit address for the fixed amount. Any BTC sent to this address becomes locked in a multi-party BitVMX setup. The functionary set is in charge to assign a different address to each one of the users.
    
    \item The user generates and signs a deposit transaction to send their BTC to the provided address.
    
    \item All functionaries collectively sign the deposit transaction. Since the input UTXOs belong to the user, this collective signing does not affect their spending. The collective signature is crucial as it serves to validate the deposit as a legitimate peg-in transaction on Rootstock. This requirement ensures that the peg-in process also operates under the 1-of-$n$ honest assumption, as all functioanries must approve the deposit.
    
    \item Once the deposit transaction is broadcast, and receives a required number of confirmations, any participant (functionary, user, or external observer) can inform Rootstock about it. 
    
    \item Rootstock then validates the transaction's inclusion in Bitcoin and the functionaries' signatures, after which it releases the corresponding amount of wrapped BTC to the user's address on Rootstock. This process is guaranteed by one of the protocol's core assumptions: Rootstock's ability to validate Bitcoin transaction inclusion.
\end{enumerate}  

\subsection{Peg-out process}

The peg-out process, which enables users to transfer wrapped BTC from Rootstock back to Bitcoin, operates as follows:

\begin{enumerate}
    \item The user creates a peg-out transaction in Rootstock, sending their wrapped BTC to a special peg-out address. Peg-out amounts are restricted to the same set of predetermined values as peg-ins. 
   
    \item The peg-out transaction is then uniquely linked to one of the UTXOs from a previous peg-in secured by a BitVMX setup. This linking can be handled either by the functionary set or automatically by Rootstock.
    
    \item After a required number of confirmations in Rootstock, a functionary acting as the operator advances (or fronts) the BTC to the user on Bitcoin. As discussed later, the amount fronted to the user may be reduced as a fee. At this point, the user has effectively completed their peg-out.
    
    \item After a number of confirmations on Bitcoin, the operator submits the Bitcoin transaction that advanced the funds to the user to Rootstock for validation. This generates a proof of peg-out on Rootstock.

    \item In order to recover the advanced funds from the linked BitVMX setup, the operator submits a \textit{Kick-off} transaction with a proof of a valid execution of \texttt{checkChain} (see Section~\ref{sec:validation}).
    
    \item The remaining functionaries act as verifiers and can challenge the proof - if they believe it is invalid. If no challenges are raised, the operator can access the funds after a time lock period, completing the peg-out process.
\end{enumerate}

\subsection{Security deposits}

The protocol mandates that functionaries maintain security deposits to ensure proper behavior when acting as operators or verifiers.
These deposits are held in Bitcoin, although could also be held on Rootstock (or a generic secondary blockchain system).
These deposits serve multiple purposes. 
They incentivize honest behavior by making malicious actions economically costly. 
In addition, they allow honest functionaries to recover the costs incurred while proving their correct assertions when challenged to do so.
For functionaries who act honestly, the only cost of operating the bridge should be the opportunity cost of their locked security deposits.
We can distinguish between two types of costs in operating the bridge. 
First, there are the operational costs, which are the regular expenses incurred during normal bridge operations. 
These include transaction fees for standard peg-in and peg-out operations, infrastructure and maintenance costs, and administrative overhead. 
These costs should be covered by the fees paid by users for using the bridge services.
Second, there are the security enforcement costs, which arise when dealing with malicious behavior. 
These include expenses related to executing the challenge-response protocol, conducting dispute resolution transactions, and performing emergency protocol operations. 
These costs should be covered by slashing the security deposits of the malicious actors involved.
For the purpose of calculating security deposits, we assume that operational costs are covered by user fees, and focus on ensuring that security deposits are sufficient to cover all potential security enforcement costs.

To achieve these numerous benefits, the total amount of locked funds required can be substantial.

To address the potentially high costs of security deposits while maintaining the protocol's security properties, we introduce two key concepts that improve capital efficiency: 

\begin{itemize}
\item \emph{Packets:} Collections of VMXOs that share the same security deposits, allowing multiple peg-out operations to be secured by a fixed set of security deposits.
  \item \emph{Enablers:} Special UTXOs that act as participation tokens, required for claiming funds from a peg-out (i.e from a VMXO) or for claiming rewards for winning challenges.
\end{itemize}

These mechanisms work together to optimize the use of security deposits while preserving the protocol's security guarantees.
To effectively manage multiple peg operations sharing the same security deposit, the protocol requires a mechanism to exclude malicious actors. 
When a functionary acts maliciously in any peg operation associated with a shared security deposit, two outcomes must be guaranteed: their security deposit is slashed, and they are prevented from participating in any other operations secured by that deposit. 
This ensures that a single instance of malicious behavior results in complete removal from all related protocol operations.

\subsubsection{Packets}

To maintain capital efficiency, the protocol uses the notion of packets. A packet is a collection of VMXOs that:
\begin{itemize}
    \item Share the same security deposits
    \item Are initialized simultaneously during the setup phase
    \item Share the same set of functionaries
\end{itemize}

This packet-based approach improves capital efficiency by allowing multiple peg operations to be secured by the same deposits, rather than requiring individual security deposits for each BitVMX setup (VMXO) or peg operation.

A key feature of the packet system is that, within a packet, each functionary can serve as operator for at most one active peg-out process at a time.
This intentional constraint helps to cap the maximum required security deposit. 
The Union bridge is designed to safeguard honest functionaries, while any additional costs due to misbehavior are borne by the malicious party. 
Therefore, each security bond must cover:
\begin{itemize}
    \item The cost of executing a single peg-out as an operator (prover)
    \item The cost of challenging up to $N-1$ concurrent peg-outs as a verifier, where $N$ is the total number of functionaries (verifier)
\end{itemize}

In this way, security deposits are both predictable and manageable, while also compelling any malicious party to bear the additional costs they cause.
The deposit is locked until either all peg-out operations in a packet are successfully completed or the functionary renounces their role for the remaining peg-outs.
In case of misbehavior, a functionary's security deposit is slashed.

\subsubsection{Enablers}
An enabler is a specifically-crafted UTXO that must be included as an input in the \textit{Unlocking} transaction when serving as an operator. When a functionary serves as a verifier, an enabler must be used as an input in the winning transaction that slashes the security bond of the losing party. 
Enablers are specific to each functionary, to each role (operator or verifier), and each VMXO. They serve as participation tokens. If a functionary loses a dispute, then all of their remaining enablers are ``burnt'', ensuring that they can no longer participate in the protocol as an operator or verifier - for a given packet.
For each VMXO, a functionary receives one enabler to act as an operator and $N-1$ enablers to act as a verifier (where $N$ is the total number of functionaries). 
This mechanism enforces that functionaries who have had their enablers consumed (either through normal protocol operation or as a penalty) are effectively removed from participating in future protocol actions.

As depicted earlier in Figure~\ref{fig:dispute}, any dispute concludes with either a ``prover loses'' or    ``verifier loses'' transaction. 
These transactions are standardized, meaning they follow a consistent format. 
In these transactions, the winning party gains the ability to burn all the enablers from the losing party. 
Additionally, the winner earns the right to claim the corresponding security deposit from the losing party.
This is how we use the enablers to slash the security deposit of the losing party and ensure that the malicious party is not able to participate in any other peg-out operations.

This mechanism ensures that any functionary engaging in malicious behavior faces significant penalties, thereby incentivizing honest participation. 
The standardized nature of these transactions simplifies the enforcement of penalties and the transfer of security deposits, maintaining the protocol's integrity and security.

This structure enables automatic enforcement of participation restrictions and deposit slashing when misbehavior is detected through the dispute resolution process. 
This system ensures that:

\begin{itemize}
    \item If a functionary withdraws their security deposit, they are prevented from participating in any of the setups in that particular set, either as operator or verifier
    \item If a functionary is penalized through the dispute resolution process, they are similarly prevented from participating in any setups in that set
    \item The dispute resolution process can slash a functionary's deposit if they lose a dispute, whether they were acting as an operator or verifier
    \item Since the locking amount for each setup is predefined and functionaries cannot operate multiple peg-outs simultaneously, the security deposit only needs to be proportional to the funds locked in a single VMXO rather than the cumulative value of all VMXOs in a packet.
\end{itemize}

\begin{figure}[h]
  \centering
  \includegraphics[width=\textwidth]{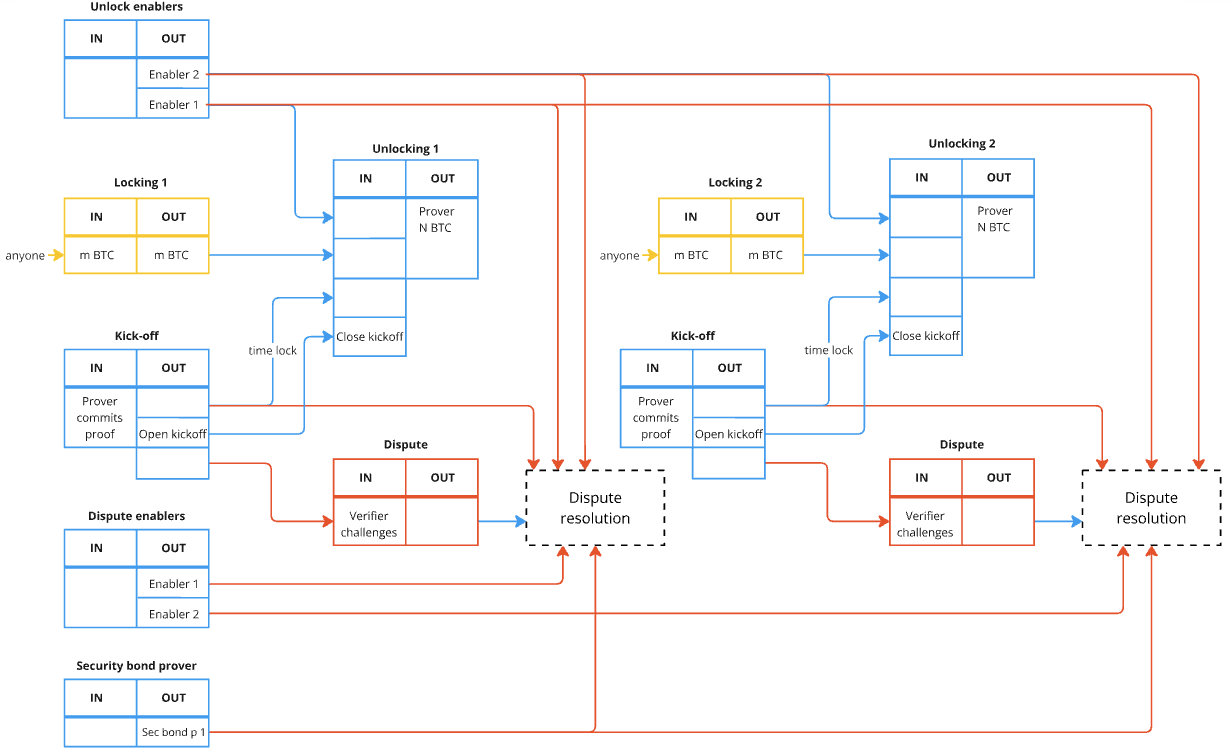}
  \caption{Security deposit management in the BitVMX proving system. The system uses shared deposits per packet and enabler outputs to manage functionary participation and enforce penalties.}
  \label{fig:deposits}
\end{figure}

\begin{figure}[h]
  \centering
  \includegraphics[width=0.8\textwidth]{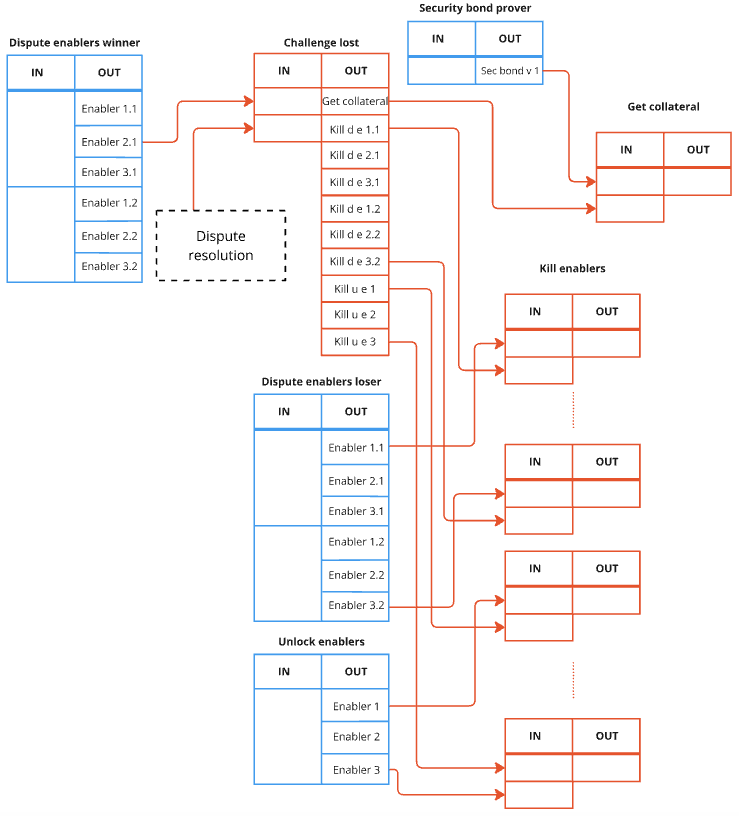}
  \caption{Security deposit management in the BitVMX challenge system. The system employs a special transaction known as \textit{Kill enablers}, which enables functionaries to prevent a malicious party from participating in future prover or challenge operations.} 
  \label{fig:enablers}
\end{figure}

The implementation of these security mechanisms is achieved through specially crafted enabler outputs in the \textit{Unlocking} transactions and the dispute resolution final transactions.
Figure~\ref{fig:deposits} illustrates the sharing of a single security bond across two distinct peg operations. 
The figure also highlights the role of enablers, which are essential for completing the \textit{Unlocking} transaction. 
The special output \textit{Open kick-off} ensures that an initiated kick-off remains locked until it is utilized in the \textit{Close kick-off}.
Each enabler is associated with a specific transaction type that can be triggered when a losing transaction is published on-chain. 
The publication of the \textit{Challenge lost} transaction allows the verifier to deactivate all remaining enablers as seen on Figure~\ref{fig:enablers}.
Additionally, this transaction releases the security bond, enabling the winning party to claim it.

\begin{figure}[h]
  \centering
  \includegraphics[width=0.6\textwidth]{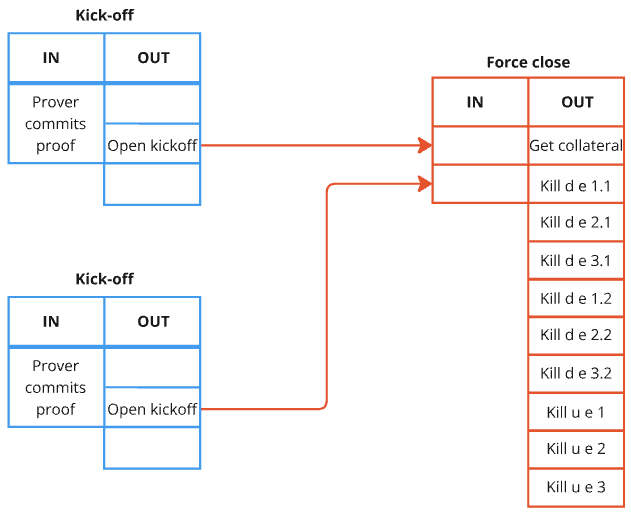}
  \caption{Force close transaction in the BitVMX system. This transaction type allows functionaries to unilaterally close a setup and reclaim their security deposit if certain conditions are met, such as prolonged inactivity or failure to reach consensus.}
  \label{fig:force_close}
\end{figure}

In order to enforce the limitation of one simultaneous peg-out process per operator, we introduce an additional transaction type called \textit{Force close}, shown in Figure~\ref{fig:force_close}. 
This transaction is created for each pair of \textit{Open kick-off} outputs generated in the \textit{Kick-off} transaction within each packet. 
The \textit{Force close} transaction enables functioanries to independently terminate a setup and recover the security deposit from a malicious operator.
As the \textit{Open kick-off} outputs are utilized in the \textit{Unlocking} transaction, any prover will consume them upon claiming the funds, completing the peg-out.

\subsection{Setup phase}

The BitVMX protocol operates through a series of co-signed transactions, the generation of which can be resource-intensive in terms of both processing time and computational power. 
To optimize this process, each transaction group is assigned a specific time limit for signature collection. 
Once the required signatures are gathered, the bridge can begin its operations without waiting for additional signatures.

A critical consideration at this stage is the cascading effect of transaction modifications: any change to an output or input of a transaction invalidates all signatures of subsequent transactions that depend on it. 
This is because transaction signatures are computed over the entire transaction data, including references to previous transactions (that are propagated through the transaction id's).
The transaction structure shown in Figure~\ref{fig:deposits} has several important implications for the protocol:

\begin{itemize}
    \item The \textit{Unlocking} transaction cannot be signed until its corresponding \textit{Locking} transaction is signed, as it depends on its outputs
    \item Any penalization path must burn all previous and subsequent enablers, requiring all enablers to be created before the first penalization transaction can be established
    \item The \textit{Kick-off} transaction must know its corresponding dispute address in advance to properly reference the relevant outputs
\end{itemize}

These dependencies introduce certain constraints on the setup process.
When a new packet of BitVMX setups is created, the process follows these steps:

\begin{enumerate}
    \item Each one of the functionaries creates a set of enablers for the packet. The exact amount will depend on the minimum amount of peg-outs and the number of operators.

    \item For each one of the peg-ins:
    \begin{enumerate}
      \item All functionaries collectively generate and sign the set of transactions required for the peg-out that do not require to know the peg-in address: kick-off, and dispute resolution transactions. This is the most expensive part since it contains all the challenge-response protocols.
    
      \item Each functionary maintains a set of private keys associated with the peg-in.
      
      \item When a peg-in is initiated, the corresponding \textit{Open kick-off} output and \textit{Locking}/\textit{Unlocking} transactions are signed. All functionaries must immediately and permanently delete their private keys associated with that specific peg-out.
      
      This key deletion is the critical security moment - if at least one honest functionary deletes their keys as required, the protocol's security is guaranteed since the only valid spending paths become those defined by the BitVMX setup.

    \end{enumerate}
\end{enumerate}

The security of the entire protocol relies on the key management process. 
As long as one functionary honestly deletes their keys after signing the \textit{Unlocking} transaction, the pegged-in funds in the VMXO can only move through the predefined BitVMX paths, regardless of the behavior of other functionaries. 
This aligns with the protocol's 1-of-$n$ honest assumption while ensuring that the locked funds remain secure through cryptographic rather than trust-based guarantees.
The protocol's design allows for gradual generation of challenge-response protocols after the initial enabler creation. 
This is a significant optimization, as bridge operations can begin as soon as enablers are created, without waiting for all challenge-response setups to be signed. 
The challenge-response protocols can be generated incrementally, spreading the computational load over time instead of concentrating it at the initialization phase. 
This method significantly shortens the interval between the creation of enablers and the commencement of bridge operations, while still upholding the protocol's security guarantees through the enabler system.
Moreover, the peg-in transaction does not need to be known beforehand to generate the subsequent challenge-response protocols.
Therefore, if all necessary transactions are prepared before the peg-in transaction is known, the peg-in transaction can be completed quickly after signing the \textit{Locking} and \textit{Unlocking} transactions.
At this stage, the primary risk is that every functionary involved in a given packet must be online and ready to sign the transactions.


\subsection{Bitcoin stop watch}

One of the key challenges in the bridge design is managing time-locks between different protocol components. 
Time-locks serve two critical purposes: they ensure parties must respond with correct answers within a reasonable timeframe, while also protecting against potential censorship attacks. 
However, the current design, which uses static time-locks in chained transactions, presents efficiency challenges.

The main inefficiency arises from these time-locks being applied sequentially, causing unnecessary delays in the protocol execution. 
For example, if each step in a chain of transactions requires a time-lock of T units of time, and there are $M$ steps, the total protocol execution could take up to $M \cdot T$ units of time.
This significantly increases the time required to execute a peg-out. 
As we limit the number of simultaneous peg-outs, this time constraint reduces the number of peg operations that can be performed within a given period.

A more efficient approach is to implement a ``stop watch'' mechanism that considers the total desired censorship resistance time as an upper bound.
We assume here eventual delivery as stated in section~\ref{sec:model}.
This time will not be applied between each one of the transactions.
Then, the waiting time has to be counted as an accumulated amount of time.
Instead of sequential time-locks, this mechanism tracks the total time a party has been waiting for the counterparty response across all protocol steps. Hence, the time for a certain party runs while it's their turn to publish a message but they do not do it.
When a party publishes a message, the watch gets started.
Afterwards, when the counterparty published the response, the watch gets stopped.
This allows us to sum all the accumulated times at the end.

\begin{figure}[h]
  \centering
  \includegraphics[width=0.9\textwidth]{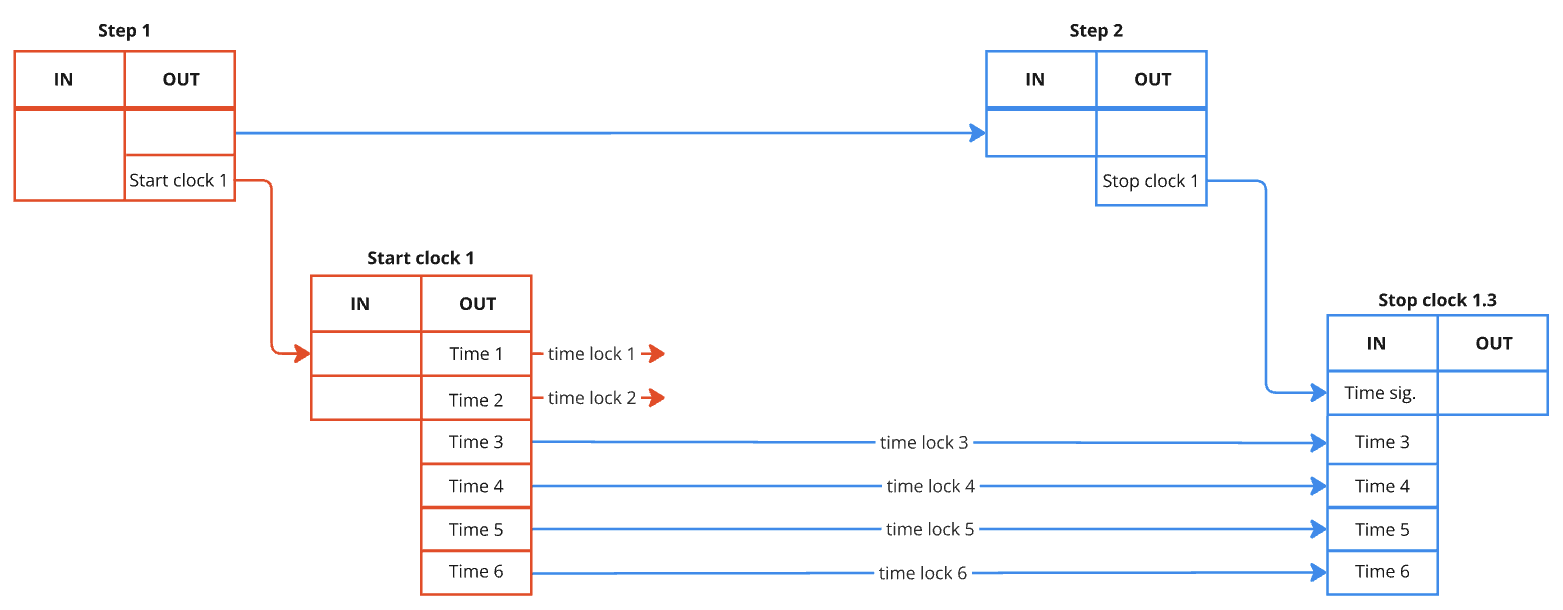}
  \caption{Time measurement between two transactions emitted by different parties.}
  \label{fig:stop_watch}
\end{figure}

When the first party $P_1$ publishes their initial transaction, it starts the watch for the counterparty $P_2$. 
$P_1$ is then responsible for mining outputs at specific time intervals (Time X) as they elapse. 
This outputs have a time-lock so they can only be mined when a certain time has elapsed. 
Once they are published, it means that this amount of time has been spent in this interval. 
As time progresses, any outputs with time-locks less than the elapsed time must be spent by $P_1$.

Once $P_2$ publishes their next transaction, they can issue a stop transaction. 
The system uses a series of stop transactions, each with one fewer slot than the previous one. 
For instance, in Figure~\ref{fig:stop_watch}, "Stop watch 1.X" signifies that $P_2$ took 2 time units to respond in the first step of the protocol. 
This elapsed time is recorded in the stop transaction by $P_2$ using Winternitz signatures in the first input. 
At this stage, the elapsed time for each interval is documented. 
This one-time signature scheme enables both parties to verify and track the total elapsed time throughout the protocol execution, as each time measurement is cryptographically committed when the subsequent transaction is published.

This mechanism ensures accurate time tracking while maintaining the protocol's security properties and allowing for efficient progression through the dispute resolution process achieving the following properties:

\begin{itemize}
  \item The total amount of time spent resolving a challenge-response protocol initiated by a verifier cannot exceed the maximum censorship resistance time
  \item Parties retain protection against censorship throughout the entire process
\end{itemize}

By signing all elapsed time with one-time signatures, we can reuse these values in subsequent transactions. 
Specifically, we can create a special transaction that aggregates all the elapsed time from previous transactions.
If this aggregated time exceeds a predetermined threshold (indicating the timeout period of the challenge-response game), the script permits the transaction to be executed, thereby burning all the functionary's enablers. 
This method provides a more efficient way to enforce protocol timeouts while preserving the security properties of the enabler system.

\section{Analysis}
\label{sec:analysis}

This section examines key design decisions and trade-offs in the Union protocol. 
We analyze the implications of functionary set size, security deposit calculations, capital efficiency through the packet system, light client flexibility, parallel operation limits, and optimization strategies for the challenge-response mechanism. 
Each aspect is evaluated in terms of its impact on security, efficiency, and practical implementation considerations.

\subsection{Functionary set size}

The size of the functionary set presents an important trade-off in the protocol's design. While a larger set of functionaries enhances the protocol's security by increasing the likelihood that at least one honest participant exists, it also impacts operational efficiency.

The protocol requires all functionaries to be connected and available at some point between the peg-in demand and the peg-in process, as their signatures are necessary to validate deposits. 
As the number of functionaries increases, so does the probability that one or more may be temporarily unavailable, potentially preventing new peg-ins from being added to a packet.
We increase capital efficiency by sharing the deposit. 
If the deposit can no longer be shared because a functionary disappears, we cannot increase the capital efficiency.
Thus, this reduced availability directly impacts the capital efficiency of packets, as they may not reach their full capacity for peg-in operations.

As previously mentioned, there are operational costs in serving in the functionary set. For sustainability, the functionaries must be able to recover these costs from user fees. As the size of the functionary set increases, bridge fees will need to rise as well. On the other hand, if a particular implementation of the Union protocol has a specific fee  target in mind (such as 0.1 percent of value bridged),  then that choice will implicitly determine the upper bounds on the size of functionary set membership.

\subsection{Security deposit calculation} \label{sec:security_bond_calculation}

If the Union bridge aims to cover all the expenses generated by the malicious party, the security deposit amount must be carefully calculated to cover the worst-case scenarios for both operators and verifiers. 
In the base scenario, the maximum number of simultaneous peg-outs is limited by the size of the functionary set ($N$), with each functionary acting as a prover for one peg-out operation.

Another important consideration is that all honest functionaries must initiate challenges against suspected malicious peg-outs independently. 
This is because:

\begin{itemize}
  \item The outcome of any single challenge-response protocol is uncertain until completion
  \item Waiting for other functionaries' challenges to conclude could leave insufficient time to initiate a new challenge
  \item If a functionary completely trusts another functionary to handle the challenge-response protocol and defeat the malicious party, there is no incentive for it to participate as a functionary itself.
\end{itemize}

When calculating the required security deposit, we must consider two scenarios:

\begin{itemize}
    \item \emph{As a prover:} A functionary must have sufficient bond to cover $N-1$ challenge-response protocols, as all other functionaries may need to challenge their proof if it is malicious.
   
     \item \emph{As a verifier:} A functionary must have sufficient bond to cover challenging up to $N-1$ concurrent valid peg-outs from other functionaries.
\end{itemize}

In addition, only the first successful challenge will trigger the slash condition. 
Therefore, the security bond must be distributed among all potential challengers to ensure coverage when multiple functionaries simultaneously challenge the same malicious peg-out. 
Specifically, the worst-case scenario occurs when a single malicious actor ready to spend funds with the only goal of disruption initiates a bad peg-out (as stated previously, all functionaries must challenge) while all other functionaries are performing legitimate peg-outs. 
At the same time, this functionary challenges the ($N-1$) correct peg-outs from the other functionaries.
This adds up to $2 \cdot (N-1)$ challenge-response protocols covered by a single security deposit.
At this stage, each functionary is involved in two challenge-response protocols with the malicious actor: one to challenge the malicious peg-out and another to execute their own legitimate peg-out.
Nevertheless, there is an appreciation. Each functionary only needs to proceed with one challenge-response with this malicious actor. 
Winning this challenge will burn all enablers, preventing the malicious actor from successfully completing any other peg-out as prover or challenge-response protocol as verifier.
Therefore, there are $(N - 1)$ malicious challenge-response protocols that need to be secured by the security deposit of a single functionary.

Considering that in the worst-case scenario a single functionary must secure $(N - 1)$ challenge-response protocols, we analyze the minimum requirements for the security deposits.
The cost of the challenge-response protocol depends on the specific operation being challenged. To determine the most expensive path in the challenge-response Directed Acyclic Graph (DAG), we must consider two factors: the amount of information shared up to the challenge transaction and the size of the challenge transaction itself. 
As detailed in the BitVMX paper~\cite{lerner24}, the longest path in the DAG occurs during a second n-ary search to resolve read values after the initial search. 
The most resource-intensive script is the one responsible for computing the SHA256 hash. 
Consequently, the SHA256 challenge at the end of the read search represents the most expensive scenario in the entire challenge-response protocol.
The exact cost of the most expensive challenge-response protocol must be calculated based on several parameters:

\begin{itemize}
    \item The number of intervals used in the n-ary search (typically 2, 4, or 8)
    \item The maximum number of computation steps allowed
    \item The cost of performing SHA256 calculations on-chain
\end{itemize}

While future optimizations may reduce these costs, the current design must account for this worst-case scenario when determining security deposits and economic parameters.

We consider here the case for 4-ary search and 16 step searches, corresponding to a maximum amount of steps supported by the current design. Table~\ref{tab:challenge_cost} contains a summary of the vBytes~\cite{wiki:vByte} used by each type of transaction. Considering the previous explanation, we can conclude that the total amount of vBytes is equal to $2513 + 653 + 5118 \cdot (16 + 15) + 205 \cdot (16 + 16) + 3105 + 1063 + 97780 = 270332$ vBytes.

\begin{table}[h]
\centering
\begin{tabular}{lr}
\toprule
Transaction Type & vBytes \\
\midrule
Commit proof & 2513 \\
Challenge & 653 \\
Publish hashes (per step) & 5118 \\
Publish step choice (per step) & 205 \\
Publish full trace & 3105 \\
Publish read trace & 1063 \\
SHA256 computation & 97780 \\
\bottomrule
\end{tabular}
\caption{Transaction costs in vBytes for each type of transaction in the challenge-response protocol.}
\label{tab:challenge_cost}
\end{table}

The cost of a challenge-response protocol is calculated as $270332 \cdot X \cdot (N-1)$, where $N$ represents the number of functionaries and $X$ is the cost in sats per vByte. 
Table~\ref{tab:security_deposit_needed} provides calculations based on different values of these parameters. 
It is shown that the required security deposit is highly sensitive to the number of functionaries, representing the most critical parameter in the Union protocol configuration. 
The required deposit ranges from approximately 0.12\bitcoinA~with 10 functionaries at low fee rates to over 8\bitcoinA~with 100 functionaries at high fee rates. 
This significant variation underscores the importance of carefully balancing functionary set size against capital requirements when deploying the protocol.

\begin{table}[h]
  \centering
  \begin{tabular}{r|r|r}
  \toprule
  Functionaries & Fee rate (sats/vByte) & Total amount (BTC) \\
  \midrule
  10 & 5 & 0.1216494 \\
  10 & 10 & 0.2432988 \\
  10 & 20 & 0.4865976 \\
  10 & 30 & 0.7298964 \\
  \midrule
  25 & 5 & 0.3243984 \\
  25 & 10 & 0.6487968 \\
  25 & 20 & 1.2975936 \\
  25 & 30 & 1.9463904 \\
  \midrule
  50 & 5 & 0.6623134 \\
  50 & 10 & 1.3246268 \\
  50 & 20 & 2.6492536 \\
  50 & 30 & 3.9738804 \\
  \midrule
  100 & 5 & 1.3381434 \\
  100 & 10 & 2.6762868 \\
  100 & 20 & 5.3525736 \\
  100 & 30 & 8.0288604 \\
  \bottomrule
  \end{tabular}
  \caption{Required security deposit based on number of functionaries and fee rates. Calculations assume worst-case scenario where all challenge-response protocols must be executed.}
  \label{tab:security_deposit_needed}
  \end{table}

  There are several potential optimizations that could significantly reduce these costs. 
  One promising approach would be to replace SHA256 with the Blake3 hashing algorithm, which requires fewer Bitcoin script instructions and would therefore be more cost-efficient. 
  Additionally, minimizing the number of rounds in the challenge-response protocol could yield substantial savings. 
  These and other optimizations suggest considerable room for improving the protocol's efficiency.

\subsection{Packet size and capital efficiency}

The packet system is designed to maximize capital efficiency by allowing multiple peg-out operations to share the same security deposits. This approach significantly reduces the total security deposit required to operate the bridge relative to the total value locked (TVL) or the amount of peg operations.

The security deposits are calculated based on protocol execution costs rather than the locked amounts. Specifically:

\begin{itemize}
    \item The required security deposit is proportional to potential dispute resolution costs, not the value being transferred
    \item A single security deposit can secure multiple peg-outs within the same packet
\end{itemize}

The maximum size of a packet is determined by the number of enablers created during the initial setup. 
The initial transactions (\textit{Locking}, \textit{kick-off}, and basic dispute structure) can be generated at any time between the enablers generation and the corresponding peg-in. 
Subsequent transactions (mainly the dispute resolution) for specific peg-outs can be generated afterwards, either on demand or incrementally little by little. 
This allows for huge packet sizes while maintaining security guarantees through the enabler system and having a reasonable time between the packet creation and the first peg-in.
After a peg-out operation is completed, participants may retain some unused enablers. 
For each peg-out, an equal number of enablers as there are functionaries is generated. 
These unused enablers can be utilized in future peg operations within the same packet.
Specifically, there are two scenarios to consider:
\begin{itemize}
\item If a challenge occurs, the malicious party will be removed from the packet. The enablers associated with this party, used by other functionaries, can be repurposed.
\item If no challenge arises, only the functionary acting as the operator will consume an enabler in the \textit{Unlocking} transaction. The remaining functionaries will retain their enablers from this peg-out, which can be reused in future peg-in operations.
\end{itemize}
There might be instances where all enablers are tied up in ongoing peg operations. 
However, with a balanced use of peg-ins and peg-outs, many enablers will be recycled throughout the bridge's operation.

It's crucial to highlight that although the complete initial setup generation can be costly, once the enablers are created, subsequent setups can be incrementally generated while the first peg-in is already underway. 
This approach allows us to avoid generating all transactions simultaneously, which is a significant time-optimizing measure.

This design achieves superior capital efficiency compared to systems requiring individual deposits for each operation, as the security deposit only needs to cover the worst-case dispute resolution scenario rather than being proportional to the total value locked in the bridge.

\subsection{Light client}

The Union protocol's design allows for flexible validation of competing proofs through the use of a metric. When two apparently valid proofs are presented the protocol can determine the legitimate proof by comparing their associated metrics as explained in Figure~\ref{fig:counter}.

The specific metric can be tailored to the particular blockchain system being bridged. For example:

\begin{itemize}
    \item In a Proof-of-Work side-chain scenario, the metric could be the accumulated proof-of-work in the chain preceding the proving transaction
    \item For rollups on Bitcoin, the metric might be the amount of proof-of-work containing the transaction being verified
    \item Other consensus mechanisms could define their own appropriate metrics based on their security models
\end{itemize}

This flexible approach to proof validation makes the protocol adaptable to various blockchain architectures and consensus mechanisms. Rather than being tied to a specific validation scheme, the protocol can accommodate different metrics while maintaining its core security properties. This generality allows the Union protocol to serve as a bridge framework that can be implemented across diverse blockchain ecosystems.

\subsection{Maximum amount of parallelism}

While the base protocol design limits each functionary to one active peg-out operation per packet at a time, this constraint can be relaxed through careful time management. Multiple parallel peg-outs can be allowed if sufficient time separation exists between them, ensuring that malicious behavior can be detected and stopped before subsequent operations complete.

Consider a scenario where a functionary initiates multiple peg-outs in sequence:

\begin{itemize}
    \item Let $T_{max}$ be the maximum time required to complete a challenge-response protocol
    \item Let $T_{min}$ be the minimum time required to complete a challenge-response protocol
    \item Let $T_f$ be the time required to execute a \textit{Force close} transaction
    \item Let $T_s$ be the safety margin to account for network delays and block time variance
\end{itemize}

If we enforce a minimum time separation of $T_{sep} = (T_{max}-T_{min}) + T_f + T_s$ between consecutive peg-outs from the same functionary, then:

\begin{itemize}
    \item Any malicious behavior in the first peg-out can be detected and challenged
    \item The \textit{Force close} transaction can be executed within $T_f$
    \item The safety margin $T_s$ ensures the \textit{Force close} completes before the next peg-out reaches a critical state
\end{itemize}

This time-based separation means that honest functionaries only need to challenge the first malicious peg-out in a sequence. Once that challenge succeeds and the \textit{Force close} executes, the malicious functionary's enablers are burned, preventing them from completing any subsequent peg-outs.

The maximum number of parallel peg-outs ($P_{max}$) that can be safely allowed for each functionary is therefore:

\[ P_{max} = \left\lfloor\frac{T_{total}}{T_{min}}\right\rfloor \]

Where $T_{total}$ is the total timelock period allocated for the packet's operations.

This approach maintains the protocol's security guarantees while allowing increased parallelism, as the time separation ensures that security deposits only need to cover challenging a single malicious peg-out rather than all concurrent operations. However, implementers must carefully consider the trade-off between increased parallelism and longer total timelock periods when choosing these parameters.

This analysis must consider multiple network vulnerabilities, including potential censorship, network forks, connectivity issues, and unplanned functionary outages.

\section{Conclusions}
\label{sec:conclusion}

In this paper, we present Union, a novel trust-minimized bridge protocol that enables secure transfer of BTC between Bitcoin and Rootstock. 
Union leverages BitVMX, an optimistic proving system on Bitcoin, in a multi-party setting to create a bridge that operates securely under the assumption that at least one participant remains honest. 
The protocol introduces several key innovations:

\begin{itemize}
    \item A packet-based architecture that allows multiple bridge operations to share security deposits, significantly improving capital efficiency
    \item An enabler system that effectively manages functionary participation and enforces penalties for misbehavior
    \item A flexible light client framework that can adapt to various blockchain architectures and consensus mechanisms
    \item An efficient stop watch mechanism that optimizes time-lock management in the challenge-response protocol
\end{itemize}

Union achieves these capabilities while maintaining strong security guarantees and minimizing trust assumptions. 
The protocol is designed to be practical and scalable, with careful consideration given to operational costs, network constraints, and implementation complexity. 
Our analysis demonstrates that Union provides a robust foundation for Bitcoin interoperability that can be adapted to various secondary blockchain systems while preserving Bitcoin's security properties.

\bibliographystyle{splncs04}
\bibliography{biblio}

\end{document}